\documentclass[12pt]{article}
\usepackage{natbib}
\usepackage{graphicx}
\usepackage{amssymb,amsfonts,amsmath}
\usepackage{geometry}
\usepackage{setspace}

\begin{document}
\title{Modelling the covariance structure in marginal multivariate count models: Hunting in Bioko Island}
\author{W. H. Bonat\thanks{Department of Mathematics and Computer Science, University of Southern Denmark, Odense, Denmark.
		    Department of Statistics, Paran\'a Federal University, 
		    Centro Polit\'ecnico, Curitiba 81531980,
		    CP 19081, Paran\'a, Brazil. E-mail: wbonat@ufpr.br} 
~and J. Olivero\thanks{Universidad de M\'alaga, Grupo de Biogeograf\'ia, Diversidad y Conservaci\'on, 
		      Departamento de Biologia Animal, Facultad de Ciencias, Campus de Teatinos s/n, 
		      29071 M\'alaga, Spain.} 
~and M. Grande-Vega\thanks{Research Group for Sustainable Management Silvanet, Faculty of Forestry, 
		   Technical University of Madrid, Ciudad Universitaria, 28040 Madrid, Spain and 
		   Asociaci\'on Ecotono, Paseo de la Habana 109. 2 A 28036 Madrid, Spain.} \\
~and M. A. Farf\'an\thanks{Universidad de M\'alaga, Grupo de Biogeograf\'ia, Diversidad y Conservaci\'on, 
		   Departamento de Biolog\'ia Animal, Facultad de Ciencias, Campus de Teatinos s/n, 29071 M\'alaga, 
		   Spain and BioGea Consultores, C/Navarro Ledesma 243, Portal 4-3 C, 29010 M\'alaga, Spain.} 
~and J. E. Fa\thanks{Division of Biology and Conservation Ecology, School of Science and the Environment, Manchester 
	     Metropolitan University, Manchester M1 5GD, UK and Center for International Forestry Research (CIFOR), 
	     CIFOR Headquarters, Bogor 16115, Indonesia.}}

\maketitle

\begin{abstract}
We present a flexible statistical modelling framework to deal with multivariate count data along with longitudinal and repeated measures structures. The covariance structure for each response variable is defined in terms of a covariance link function combined with a matrix linear predictor involving known matrices. To specify the joint covariance matrix for the multivariate response vector the generalized Kronecker product is employed. The count nature of the data is taken into account by means of the power dispersion function associated with the Poisson-Tweedie distribution. Furthermore, the score information criterion is extended for selecting the components of the matrix linear predictor. We analyse a dataset consisting of prey animals (the main hunted species, the blue duiker \textit{Philantomba monticola} and other taxa) shot or snared for bushmeat by $52$ commercial hunters over a $33$-month period in Pico Basil\'e, Bioko Island, Equatorial Guinea. By taking into account the severely unbalanced repeated measures and longitudinal structures induced by the hunters and a set of potential covariates (which in turn affect the mean and covariance structures), our method can be used to indicate whether there was statistical evidence of a decline in blue duikers and other species hunted during the study period. Determining whether observed drops in the number of animals hunted are indeed true is crucial to assess whether species depletion effects are taking place in exploited areas anywhere in the world. We suggest that our method can be used to more accurately understand the trajectories of animals hunted for commercial or subsistence purposes, and establish clear policies to ensure sustainable hunting practices.
\end{abstract}


\section{Introduction}

Multivariate regression models have been of increased interest in the 
statistical literature. Recent applications include functional 
disability data \citep{Vallier:2014}, cognitive functioning 
\citep{Anderlucci:2015}, evolutionary biology \citep{Cybis:2015}, 
multi-species distribution \citep{Hui:2015,Ovaskainen:2011},
social, economic \citep{Klein:2015,Klein:2015a} and political sciences 
\citep{Lagona:2015} to cite a few.

The mentioned methodologies apply latent variables or finite mixture of
regression models to describe the covariance structure introduced by the
multiple response variables. 
In contrast to these approaches \citet{Bonat:2016} proposed the 
multivariate covariance generalized linear models (McGLMs), which 
explicitly model the marginal covariance matrix combining a covariance 
link function and a matrix linear predictor composed of known matrices.
McGLMs have much in common with the GEE (Generalized Estimating Equations)
\citep{Zeger:1988} approach popular in the analysis of longitudinal data.
However, McGLMs were explicitly designed to deal with multiple response
variables and allow for a flexible modelling of the covariance structure.
On the other hand, current GEE implementations~\citep{Hojsgaard:2006} 
deal only with one response variable and include a short list of pre-specified
covariance structures, such as autoregression and compound symmetry.

Generalized linear mixed models (GLMMs) \citep{Breslow:1993} are 
flexible models for handling multivariate data 
\citep{Verbeke:2014}. GLMMs are computationally demanding, and many 
algorithms have been proposed in the past three decades, see 
\citet{McCullogh:1997} and \citet{Fong:2010} for reviews and further
references. \citet{Motta:2013} presented a specific example of
GLMM for count data.
An aspect of GLMMs that gives rise to concern is the general
lack of a closed-form expression for the likelihood and the marginal
distribution of the data vector. A related question is the special 
interpretation of parameters inherent from the construction of GLMMs.
Thus, the covariate effects are conditional on the latent variables,
whereas the correlation structure is marginal for the latent variables
rather than for the response variables.

The multivariate Poisson \citep{Tsionas:1999} and negative binomial 
\citep{Shi:2014} distributions are suitable approaches to deal with 
multivariate count data. The multivariate Poisson has the restriction 
to deal only with equidispersed and positive correlated data. 
The last restriction is also shared by the multivariate negative 
binomial model. The assumption of a common error distribution required 
for these models may not be satisfied in practice, and methods for 
handling the case of unequal marginal distributions do not seem easily 
available.  Additional methods for specifying 
models for dependent data include the Gaussian copula marginal regression 
models \citep{Masarotto:2012} and the class of hierarchical 
generalized linear models \citep{Lee:1996}.

In the context of multivariate longitudinal models, besides the 
modelling of the covariance structure between response 
variables, we also have to model the longitudinal and repeated measures 
structures for each response variable, i.e. the within covariance structure.
The question of how to model the within covariance structure in the 
univariate case is often solved by choosing from a short list of options, 
such as compound symmetry, autoregressive and 
unstructured~\citep{Diggle:2002}.
Such choices are, however, not suitable for the combination of multivariate, 
repeated measures and longitudinal structures found in the application 
described in the Section~\ref{dataset}. 
It motivates the development of a more general and flexible approach 
for covariance modelling in multivariate longitudinal count models.

In this paper, we adopt the McGLM framework in order to present 
a multivariate model suitable to deal with count response variables. 
Our model also relies on the structure of the 
multivariate discrete dispersion models~\citep{Jorgensen:2014}, where 
the Poisson-Tweedie distribution provides a flexible framework for 
modelling discrete response variables. In this framework multivariate 
extensions of the Neyman Type A, P{\'o}lia-Aepply, negative binomial and
Poisson-inverse Gaussian distributions appear as special cases. 
One advantage of this class of models is that similar to the exponential
dispersion models \citep{Jorgensen1997b} the whole family
is described by the power dispersion function, analogous to ordinary 
Tweedie exponential dispersion models with power variance functions. 
This fact allows us to specify models based on second-moment assumptions
and use the engine of McGLMs for estimation and inference.
For further references and regression models based on the Poisson-Tweedie
distribution, see~\citet{Bonat:2016c}.

The model is motivated by a data set consisting of the number of blue 
duikers and other small animals shot or snared by $52$ commercial 
hunters over a $33$-month period in Pico Basil{\'e}, Bioko Island, 
Equatorial Guinea \citep{Vega:2015}. Bushmeat trade is an important 
resource in the livelihoods of many rural communities in West and 
central Africa. Overhunting for profit is known to cause immediate 
reductions in the density of targeted animals \citep{Fa:2000}. 
In extreme cases it may precipitate the disappearance of local 
populations and eventually result in the complete extirpation of a 
species \citep{Fa:2009}. It is also known that hunted island animal 
populations are often at a greater risk of extinction because of their 
small geographic ranges and usually low population numbers 
\citep{Vega:2015}. In Bioko Island, the blue duiker 
(\textit{Philantomba monticola}) is the most hunted species among $18$ 
species of mammals and birds consumed as food.

The main goal of this data analysis is to investigate whether the number of hunted 
blue duikers declined during the study period. The data analysis should 
take into account the severely unbalanced repeated measures and longitudinal 
structures introduced by the hunters and a set of potential covariates 
affecting both the mean and covariance structures. 

Determining whether the decline of hunted animals is instrumental, 
since it could suggest a reduction in the population of this species, 
with important applications for establishing policies of sustainable 
hunting practices. In this scenario, a bivariate count model 
is useful, since a significant negative correlation could indicate that 
hunters target another species as a result of the decline in the target 
species, while a non-significant correlation may push hunter to 
turn to alternative sources of income.

In view of the recent developments in the McGLMs framework the main contributions
of this article are: i) introduces a suitable specification of the McGLMs 
to deal with the combination of longitudinal and repeated measures in the context 
of multivariate count data. ii) describes how to specify the components of the
matrix linear predictor in order to take into account the effects of known covariates
in a linear mixed model fashion. iii) extends the score information criterion (SIC)
to select the components of the matrix linear predictor. iv) applied the methods
to analyse the Hunting data set and v) provides \texttt{R} code for constructing the
components of the matrix linear predictor as well as fitting the models through the 
\texttt{mcglm}~\citep{Bonat:2016a} package for the \texttt{R} statistical software.

We present the Hunting data set in Section~\ref{dataset}. 
Section~\ref{model} discusses the model and its properties.
We emphasize the specification of the matrix linear
predictor. Section~\ref{sic} extends the score information criterion 
for selecting the components of the matrix linear predictor. 
Section~\ref{results} describes the application of the model to the data. 
Section~\ref{discussion} discusses the main results.
Finally, Section~\ref{conclusion} presents the concluding remarks.
The data set that is analysed in the paper and the programs that were 
used to analyse it can be obtained from\\ 
\texttt{http://www.leg.ufpr.br/doku.php/publications:papercompanions:hunting\\bioko2016}.

\section{Data set} \label{dataset}

The case study analysed in this paper uses data of animals hunted in 
the village of Basil{\'e} Fang, Bioko Norte Province, Bioko Island, 
Equatorial Guinea. The monthly number of blue duikers and other small 
animals shot or snared were collected from a random sample of $52$ 
commercial hunters from August $2010$ to September $2013$. 
For each animal caught, the species, sex, method of capture and 
altitude were documented. The data set has $1216$ observations.
For additional description of the field work, see \citet{Vega:2015}.

In this analysis, we opted to aggregate the species into two levels blue 
duikers (\texttt{BD}) and other small animals (\texttt{OT}), 
since \texttt{BD} is the target species and \texttt{OT} are hunted at 
random. The covariates \texttt{sex} (Female, Male) and 
\texttt{method} (Firearm, Snare) are factors with two levels. 
The covariate \texttt{alt} is a factor with $5$ levels 
($300\--600$, $601\--900$, $901\--1200$, $1201\--1500$ and $>1500$) 
indicating the altitude where the animal was caught. 
Finally, the number of hunter days per month was recorded. 
It is important, because represents the effort employed by the hunter 
and should be used as an \texttt{offset}(in logarithm scale) for modelling 
the counts of hunted animals. 

The study design introduces some sources of dependence in the data. 
We call \texttt{hunter-month} the effect of all observations taken at 
the same hunter and month. The \texttt{hunter} effect 
is represented by all observations taken at the same hunter. 
The \texttt{longitudinal} effect is introduced by the observations 
taken at sequentially months. The within covariance for each outcome 
can also be affected by the covariates in a linear mixed model 
fashion, see Section~\ref{model} and \citet{Demidenko:2013} for details. 
Finally, the correlation between response 
variables should be taken into account, since it plays an important role
in terms of model interpretation. 
The number of observations per \texttt{hunter-month} and \texttt{hunters}  
varied between $1$ and $16$ and $1$ and $104$, respectively. 
These numbers show the severely unbalanced repeated measures and longitudinal structures
present in the data set.

\setkeys{Gin}{width=0.99\textwidth}
\begin{figure}[h]
\centering
\includegraphics{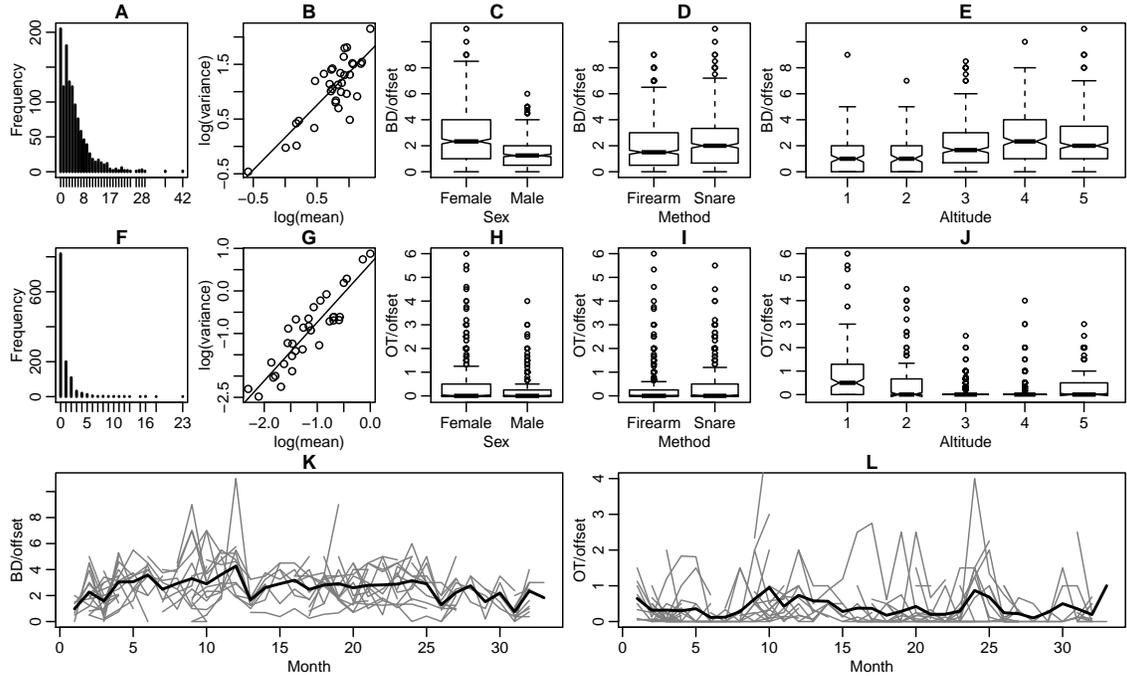}
\caption{Histograms (A and F). Taylor plot (hunter mean and variance in 
double logarithmic scale) (B and G). Boxplots for \texttt{sex} (C and H), 
\texttt{method} (D and I) and \texttt{alt} (E and J). Individual 
average (gray) and overall average (black) trajectories (K and L) for 
\texttt{BD} and \texttt{OT}, respectively.}
\label{fig:descritiva}
\end{figure}

Histograms in Figure~\ref{fig:descritiva} suggest that the two error
distributions may not be identical, and hint at potential problems with 
excess of zeroes and overdispersion. Boxplots suggest an effect of
all covariates, whereas the approximate linearity of the Taylor plots
suggest a variance function of power form.

\section{Multivariate longitudinal models for count data} \label{model}

Let $\mathbf{Y}_{N \times R} = \{\boldsymbol{Y}_1, \ldots, \boldsymbol{Y}_R\}$ 
be a response variable matrix and let 
$\mathbf{M}_{N \times R} = \{\boldsymbol{\mu}_1, \ldots, \boldsymbol{\mu}_R\}$ 
denote the corresponding matrix of expected values. 
Let $\boldsymbol{\Sigma}_r$ denote the $N \times N$ covariance matrix 
within the response variable $r$ for $r = 1, \ldots, R$. 
Similarly, let $\boldsymbol{\Sigma}_b$ be the $R \times R$ correlation matrix
whose components $\rho_{rr^{\prime}}$'s denote the correlation between the
response variables $r$ and $r^{\prime}$. 
The multivariate covariance generalized linear model as proposed by 
\citet{Bonat:2016} is given by
\begin{eqnarray}
\label{McGLM}
\mathrm{E}(\mathbf{Y}) &=& \mathbf{M} = \{g_1^{-1}(\boldsymbol{X}_1 \boldsymbol{\beta}_1), \ldots, g_R^{-1}(\boldsymbol{X}_R \boldsymbol{\beta}_R)\} \nonumber    \\
\mathrm{Var}(\mathbf{Y}) &=& \boldsymbol{C} = \boldsymbol{\Sigma}_R \overset{G} \otimes \boldsymbol{\Sigma}_b \nonumber
\end{eqnarray}
where $\boldsymbol{\Sigma}_R \overset{G} \otimes \boldsymbol{\Sigma}_b = 
\mathrm{Bdiag}(\tilde{\boldsymbol{\Sigma}}_1, \ldots, 
\tilde{\boldsymbol{\Sigma}}_R)(\boldsymbol{\Sigma}_b \otimes 
\boldsymbol{I})\mathrm{Bdiag}(\tilde{\boldsymbol{\Sigma}}_1^T, \ldots, 
\tilde{\boldsymbol{\Sigma}}_R^T)$ 
is the generalized Kronecker product \citep{Martinez:2013}. 
The matrix $\tilde{\boldsymbol{\Sigma}}_r$ denotes the lower triangular 
matrix of the Cholesky decomposition of $\boldsymbol{\Sigma}_r$. 
The operator $\mathrm{Bdiag}$ denotes a block diagonal matrix and 
$\boldsymbol{I}$ denotes an $R \times R$ identity matrix. 
The functions $g_r$ are link functions, for which we adopt the 
orthodox $\log$-link function. Let $\boldsymbol{X}_r$ denote an 
$N \times k_r $ design matrix and $\boldsymbol{\beta}_r$ a 
$k_r \times 1$ regression parameter vector. Note that, the model has a 
specific linear predictor for each response variable.

In order to specify the covariance within response variables, 
we adopt the definition of \citet{Jorgensen:2014} for Poisson-Tweedie 
random vector, i.e.
\begin{equation*}
\boldsymbol{\Sigma}_r = \mathrm{diag}(\boldsymbol{\mu}_r) + 
\mathrm{V}(\boldsymbol{\mu}_r;p_r)^{\frac{1}{2}} (\boldsymbol{\Omega}(\boldsymbol{\tau}_r))
\mathrm{V}(\boldsymbol{\mu}_r;p_r)^{\frac{1}{2}}
\end{equation*}
where $\mathrm{V}(\boldsymbol{\mu}_r;p_r) = \mathrm{diag}(\boldsymbol{\mu}_r^{p_r})$,
is a diagonal matrix whose main entries are given by the power variance function.
This specification is a multivariate representation of the power dispersion function which characterizes
the Poisson-Tweedie family, see \citet{Jorgensen:2014} for details.
Finally, following the ideas of \citet{Anderson:1973} and \citet{Pourahmadi:2000} 
we model the dispersion matrix  $\boldsymbol{\Omega}(\boldsymbol{\tau}_r)$ as a linear combination of known matrices, i.e.
\begin{equation}
\label{linearcovariance}
h(\boldsymbol{\Omega}(\boldsymbol{\tau}_r)) = \tau_{r0} Z_{r0} + \cdots + \tau_{rD} Z_{rD}.
\end{equation}
Here $h$ is the covariance link function, $Z_{rd}$ with $d = 0, \ldots, D$ 
are known matrices reflecting the covariance structure within the 
response variable $r$, and 
$\boldsymbol{\tau}_r = (\tau_{r0}, \ldots, \tau_{rD})$ is a $(D+1) \times 1$ 
parameter vector. This structure is a natural analogue of the linear 
predictor of the mean structure, and following \citet{Bonat:2016} 
we call it a matrix linear predictor. 

In this paper we focus on the identity covariance link function, 
since many interesting models appear as special cases. 
\citet{Demidenko:2013} showed that the covariance structure induced
by the orthodox Gaussian linear mixed model is a linear covariance matrix, 
i.e. has the form of (\ref{linearcovariance}). 
In this sense, the models presented in this paper can been seen as an 
extension of the Gaussian linear mixed model for handling count data.
Furthermore, popular approaches to deal with longitudinal autocorrelated 
data, as the compound symmetry, moving average and first order 
autoregressive, are also covariance linear models. In what follows 
we discuss some of the possibilities for the specification of the 
matrix linear predictor in the context of longitudinal data. 

Since the matrix linear predictor is specified for each response variable,
suppose without loss of generality that $r=1$. 
Denote $y_{go}$ an observation $o = 1, \ldots, O_g$ within
the group $g = 1, \ldots, G$ and let $\boldsymbol{y}_g$ denote the
$O_g$-dimensional vector of measurements from the $g$th group.
In particular, for the data set presented in Section~\ref{dataset} the groups 
are given by the \texttt{Hunters}.
Thus, the response variable vector is given by 
$\boldsymbol{Y} = (\boldsymbol{y}_1, \ldots, \boldsymbol{y}_G)^\top$.
Let $A_g$ denote an $O_g \times E$ design matrix composed of the 
values of $E$ known covariates available to model the covariance structure.
Furthermore, let $A_{g,\cdot e}$ denote the $e$th column of the matrix $A_g$.
Following \citet{Demidenko:2013} the main effect of the covariate $e$ 
and the interaction effect between the covariates $e$ and $e^{\prime}$ are
included in the covariance model through the symmetric matrices  
$$A_{g}^{e} = A_{g,\cdot e} A_{g, \cdot e}^{\top} \quad \mbox{and} \quad  
A_{g}^{e e^{\prime}} = A_{g,\cdot e} A_{g, \cdot e^{\prime}}^T + A_{g,\cdot e^{\prime}} A_{g, \cdot e}^T,$$
respectively. The matrices $A_{g}^{e}$ and $A_{g}^{e e^{\prime}}$ are group specific. 
To obtain the components of the matrix linear predictor for the 
entire response variable vector $\boldsymbol{Y}$, 
we assume independent groups. Thus, the components of the matrix
linear predictor that measure the effect of the $e$th covariate and the
interaction effect are given by 
\begin{equation}
\label{matlinear}
Z_e = \mathrm{Bdiag}(A_{1}^{e}, \ldots, A_{G}^e) \quad \mbox{and} \quad
Z_{ee^{\prime}} = \mathrm{Bdiag}(A_{1}^{ee^{\prime}}, \ldots, A_{G}^{ee^{\prime}}),
\end{equation}
where as before the operator $\mathrm{Bdiag}$ denotes a block diagonal matrix.
The matrices $Z_e$ and $Z_{ee^{\prime}}$ can be included as the $Z_{d}$'s components
in the matrix linear predictor, see~\ref{linearcovariance}. 
When the main and interaction effects are included in the model, 
we have $E(E+1)/2$ components. A simplification is obtained by considering 
only main effects resulting in $E$ components. 
In general, we reserve the first component of the matrix linear predictor 
$Z_0$ to an identity matrix, that represents the intercept of the linear covariance model.

\citet{Demidenko:2013} showed that some well known covariance structures used to model
longitudinal and repeated measures data are linear covariance models. 
To describe these structures consider a particular group $g$ with three observations.
As before to extend the matrices to the entire response variable vector, we assume
independent groups and use the $\mathrm{Bdiag}$ operator.
The compound symmetry or exchangeable structure is a linear combination of an identity and a 
matrix of ones, i.e. for this particular group the matrix linear predictor is given by
\begin{equation*}
\boldsymbol{\Omega}_g(\boldsymbol{\tau}) = \tau_{0} \begin{bmatrix}
1 & 0 & 0\\ 
0 & 1 & 0\\ 
0 & 0 & 1
\end{bmatrix} + \tau_{1} \begin{bmatrix}
1 & 1 & 1\\ 
1 & 1 & 1\\ 
1 & 1 & 1
\end{bmatrix} .
\end{equation*}
The Moving Average model of order $p$ MA(p) is also a linear covariance model. 
The components of the matrix linear predictor associated with the MA(1) and MA(2) 
structures are given respectively by
\begin{equation}
\label{compsyme}
A_1 = \begin{bmatrix}
0 & 1 & 0\\ 
1 & 0 & 1\\ 
0 & 1 & 0
\end{bmatrix} \quad \mbox{and} \quad
A_2 = \begin{bmatrix}
0 & 0 & 1\\ 
0 & 0 & 0\\ 
1 & 0 & 0
\end{bmatrix}. 
\end{equation}
For longitudinal data analysis, we can use the inverse of Euclidean distance between pairs
of observations as a component of the matrix linear predictor, for example
\begin{equation}
\label{Eucli}
A_1 = \begin{bmatrix}
0 & 1/d_{12} & 1/d_{13} \\ 
1/d_{12} & 0 & 1/d_{23} \\ 
1/d_{13} & 1/d_{23} & 0
\end{bmatrix},
\end{equation}
where $d_{ij}$ denotes the Euclidean distances between the observations at time $i$ and $j$.
By combining the simple structures described above, we have a flexible set of components
to compose the matrix linear predictor for the analysis of longitudinal data.
\citet{Demidenko:2013} also showed that the popular first-order autoregression model can 
be written as a linear covariance model, but using the inverse covariance link function.
In this paper, we do not pursue in this covariance link function.

The power parameter $p$ plays an important role in the context of multivariate Poisson-Tweedie
models, since it is an index which distinguishes between some important 
discrete distributions. Examples include the Neyman Type A ($p = 1$), 
P{\'o}lya-Aeppli ($p=1.5$), negative binomial ($p=2$) and Poisson-inverse 
Gaussian ($p=3$). The algorithm proposed by \citet{Bonat:2016} allows us to estimate
the power parameter, which works as an automatic distribution selection.

\section{The score information criterion}\label{sic}

In this section, we extend the score information criterion (SIC) proposed by 
\citet{Stoklosa:2014} for the selection of the components of the matrix
linear predictor. In order to introduce the SIC, we first present
some key components of the estimating function approach used to fit
McGLMs. The algorithm and asymptotic theory associated with the
estimating function estimators were presented by \citet{Bonat:2016} and 
implemented in the  \texttt{mcglm}~\citep{Bonat:2016a} package for the 
\texttt{R}~\citep{R:2015} statistical sofware.

The second-moment assumptions of McGLMs motivate us to divide the set of parameters into two subsets
$\boldsymbol{\theta} = (\boldsymbol{\beta}^{\top},\boldsymbol{\lambda}^{\top})^{\top}$.
In this notation $\boldsymbol{\beta} = (\boldsymbol{\beta}^{\top}_1, \ldots, \boldsymbol{\beta}^{\top}_R)^{\top}$ 
and $\boldsymbol{\lambda} = (\rho_1, \ldots, \rho_{R(R-1)/2}, p_1, \ldots, p_R, \boldsymbol{\tau}_1^\top, \ldots, \boldsymbol{\tau}_R^\top)^\top$ denote a $K \times 1$ and $Q \times 1$ vector of all regression and dispersion parameters, respectively.
Let $\mathcal{Y} = (\boldsymbol{Y}_1^\top, \ldots, \boldsymbol{Y}_R^\top)^\top$ and $\mathcal{M} = (\boldsymbol{\mu}_1^\top, \ldots, \boldsymbol{\mu}_R^\top)^\top$ denote the $NR \times 1$ stacked vector of the response variable matrix $\mathbf{Y}_{N \times R}$ and expected values matrix $\mathbf{M}_{N \times R}$ by columns, respectively.

The regression coefficients are estimated by using the orthodox quasi-score function \citep{Bonat:2016,Zeger:1988}.
The dispersion parameters are estimated based on the Pearson estimating function, defined by the components
\begin{equation*}
\label{Pearson}
\psi_{\boldsymbol{\lambda}_i}( \boldsymbol{\beta}, \boldsymbol{\lambda}) = \mathrm{tr}(W_{\boldsymbol{\lambda}_i}(\boldsymbol{r}^\top\boldsymbol{r} - \boldsymbol{C})) \quad \text{for} \quad i = 1,\ldots,Q,
\end{equation*}
where $W_{\boldsymbol{\lambda}_i} = -\partial \boldsymbol{C}^{-1} / \partial \boldsymbol{\lambda}_i$ and $\boldsymbol{r} = \mathcal{Y} - \mathcal{M}$.

Two key components of an estimating function approach are the sensitivity and variability matrices. 
The entry $(i,j)$ of the $Q \times Q$ sensitivity matrix of $\psi_{\boldsymbol{\lambda}}$ is given by,
\begin{equation*}
\label{Slambda}
\mathrm{S}_{\boldsymbol{\lambda}_{ij}} = \mathrm{E} \left (  \frac{\partial}{\partial \boldsymbol{\lambda}_i} \psi_{\boldsymbol{\lambda}_j} \right ) = -\mathrm{tr} \left (W_{\boldsymbol{\lambda}_i} \boldsymbol{C} W_{\boldsymbol{\lambda}_j} \boldsymbol{C} \right).
\end{equation*} 
Similarly, the entry $(i,j)$ of the $Q \times Q$ variability matrix of $\psi_{\boldsymbol{\lambda}}$ is given by
\begin{equation*}
\label{Vl}
\mathrm{V}_{\boldsymbol{\lambda}_{ij}} = \mathrm{Cov}(\psi_{\boldsymbol{\lambda}_i},\psi_{\boldsymbol{\lambda}_j}) = 2\mathrm{tr}(W_{\boldsymbol{\lambda}_i} \boldsymbol{C} W_{\boldsymbol{\lambda}_j} \boldsymbol{C}) + \sum_{l=1}^{NR} k^{(4)}_l (W_{\boldsymbol{\lambda}_i})_{ll} (W_{\boldsymbol{\lambda}_j})_{ll}, 
\end{equation*} 
where $k^{(4)}_l$ denotes the fourth cumulant of $\mathcal{Y}_l$. In order to keep the model based on second-moment assumptions only, 
we following \citet{Bonat:2016} use the empirical fourth cumulant. 

\citet{Stoklosa:2014} in the context of generalized estimating equations (GEE) proposed the score information criterion (SIC) to be used with forward selection algorithms in the cases where we have a large number of covariates to compose the linear predictor. 
The SIC is based on the score statistics, what becoming such criterion convenient, since it can be computed for all candidate models without actually fitting them. 

Suppose without loss of generality that $r = 1$ and fixed power parameter. 
In that case, the vector of dispersion parameters simplify to $\boldsymbol{\lambda} = \boldsymbol{\tau}$,
since we have no correlation neither power parameters.
For a given mean structure, suppose that the parameter vector $\boldsymbol{\tau}$ can be partitioned as $\boldsymbol{\tau} = (\boldsymbol{\tau}_{1}^\top, \boldsymbol{\tau}_{2}^\top)^\top$, whose dimension are $(Q - s) \times 1$ and $s \times 1$, respectively. The Pearson estimating function $\psi_{\boldsymbol{\lambda}}$ and its sensitivity and variability matrices, can also be partitioned to $\psi_{\boldsymbol{\lambda}}(\boldsymbol{\beta},\boldsymbol{\tau}) = (\psi_{\boldsymbol{\lambda}_1}(\boldsymbol{\beta},\boldsymbol{\tau}_1)^\top, \psi_{\boldsymbol{\lambda}_2}(\boldsymbol{\beta},\boldsymbol{\tau}_2)^\top)^\top$,

\begin{equation*}
\mathrm{S}_{\boldsymbol{\lambda}} = \begin{pmatrix}
\mathrm{S}_{\boldsymbol{\lambda}_{11}}  & \mathrm{S}_{ \boldsymbol{\lambda}_{12}} \\ 
\mathrm{S}_{\boldsymbol{\lambda}_{21}}  & \mathrm{S}_{\boldsymbol{\lambda}_{22}}
\end{pmatrix},
\end{equation*}
and
\begin{equation*}
\mathrm{V}_{\boldsymbol{\lambda}} = \begin{pmatrix}
\mathrm{V}_{\boldsymbol{\lambda}_{11}}  & \mathrm{V}_{ \boldsymbol{\lambda}_{12}} \\ 
\mathrm{V}_{\boldsymbol{\lambda}_{21}}  & \mathrm{V}_{\boldsymbol{\lambda}_{22}}
\end{pmatrix},
\end{equation*}
respectively. The null hypothesis $H_0$ is $\boldsymbol{\tau}_{2} = \boldsymbol{0}$. Let $\tilde{\boldsymbol{\tau}} = (\hat{\boldsymbol{\tau}}_{1}^\top, \boldsymbol{0}^\top)^\top$ be the vector of Pearson estimates under $H_0$. 
Note that, only the base model containing $\hat{\boldsymbol{\tau}}_{1}$ parameters has to be fitted. 
In practical situations, this model can contain only a simple intercept. 
The Pearson estimating function takes the form 
$$\psi_{\boldsymbol{\lambda}}(\beta,\tilde{\boldsymbol{\tau}}) = (\psi_{\boldsymbol{\lambda}_1}^\top(\boldsymbol{\beta},\tilde{\boldsymbol{\tau}}), \psi_{\boldsymbol{\lambda}_2}^\top(\boldsymbol{\beta},\tilde{\boldsymbol{\tau}}))^\top = (\boldsymbol{0}^\top, \psi_{\boldsymbol{\lambda}_2}^\top(\boldsymbol{\beta},\tilde{\boldsymbol{\tau}}))^\top.$$ The generalized score statistic is given by

\begin{equation}
\label{TU}
T_{\boldsymbol{\lambda}_2}(\boldsymbol{\beta}, \tilde{\boldsymbol{\tau}}) = \psi_{\boldsymbol{\lambda}_2}^\top(\boldsymbol{\beta},\tilde{\boldsymbol{\tau}}) \mathrm{Var}(\psi_{\boldsymbol{\lambda}_2}(\boldsymbol{\beta},\tilde{\boldsymbol{\tau}}))^{-1} \psi_{\boldsymbol{\lambda}_2}(\boldsymbol{\beta},\tilde{\boldsymbol{\tau}})
\end{equation}
where 
\begin{equation*}
\begin{aligned}
\label{V2}
\mathrm{Var}(\psi_{\boldsymbol{\lambda}_2}(\boldsymbol{\beta},\tilde{\boldsymbol{\tau}})) &= \mathrm{V}_{\boldsymbol{\lambda}_{22}} - \mathrm{S}_{\boldsymbol{\lambda}_{21}}\mathrm{S}_{\boldsymbol{\lambda}_{11}}^{-1}\mathrm{V}_{ \boldsymbol{\lambda}_{12}} - \mathrm{V}_{ \boldsymbol{\lambda}_{12}}\mathrm{S}_{\boldsymbol{\lambda}_{11}}^{-1}\mathrm{S}_{ \boldsymbol{\lambda}_{12}} \\
&  +  \mathrm{S}_{\boldsymbol{\lambda}_{21}}\mathrm{S}_{\boldsymbol{\lambda}_{11}}^{-1}\mathrm{V}_{\boldsymbol{\lambda}_{11}}\mathrm{S}_{\boldsymbol{\lambda}_{11}}^{-1}\mathrm{S}_{ \boldsymbol{\lambda}_{12}}
\end{aligned}
\end{equation*}
is the variance of the subvector $\psi_{\boldsymbol{\lambda}_2}(\boldsymbol{\beta},\tilde{\boldsymbol{\tau}})$. Under the null hypothesis, $T_{\boldsymbol{\lambda}_2}(\boldsymbol{\beta}, \tilde{\boldsymbol{\tau}})$ has a chi-square distribution with $s$ degrees of freedom. In practice, all quantities in (\ref{TU}) are evaluated at the Pearson estimates under the null hypotheses. If $H_0$ were true, then $\psi_{\boldsymbol{\lambda}_2}(\boldsymbol{\beta},\tilde{\boldsymbol{\tau}})$ that is the Pearson estimating function for $\boldsymbol{\tau}_2$ would be close to zero when evaluated under the null. Large values of $T_{\boldsymbol{\lambda}_2}(\boldsymbol{\beta}, \tilde{\boldsymbol{\tau}})$ would argue against $H_0$. The main idea behind SIC is to use (\ref{TU}) as a quadratic approximation to the log-likelihood ratio statistic. The so-call one-step SIC is defined by
\begin{equation*}
SIC^{(1)}(\boldsymbol{\beta}, \boldsymbol{\tau}) = - T_{\boldsymbol{\lambda}_2}(\boldsymbol{\beta}, \tilde{\boldsymbol{\tau}}) + \delta |\boldsymbol{\tau}|.
\end{equation*}
Note that this criterion is a function of $\tilde{\boldsymbol{\tau}}$ only, thus only the base model needs to be fitted. As point out by \citet{Stoklosa:2014} the approximation of score statistics to likelihood ratio statistics can be poor when there is a significant departure from the null model. Hence an improved approximation might calculate the score statistic in one-parameter increments, i.e.
\begin{equation*}
SIC(\boldsymbol{\beta}, \boldsymbol{\tau}) = - \sum_{s = 1}^{|\boldsymbol{\tau}_2|} \underset{\tau_{(s)} \in \boldsymbol{\tau}_2^{\backslash s-1}}{\mathrm{max}} \{ T_{\boldsymbol{\lambda}_2(s)}(\boldsymbol{\beta}, \tilde{\boldsymbol{\tau}}_{s-1}) \} + \delta |\boldsymbol{\tau}|
\end{equation*}
where $\boldsymbol{\tau}^\top_s = (\boldsymbol{\tau}^\top_{s-1}, \tau_s)$ and $\boldsymbol{\tau}_2^{\backslash s-1} = {\boldsymbol{\tau}_2 \cap \boldsymbol{\tau}^c_{s-1} }$ where $\boldsymbol{\tau}^c_{s-1}$ is the complement set of $\boldsymbol{\tau}_{s-1}$. In summary, we sequentially add new parameters selected from $\boldsymbol{\tau}_2$, these are $\tau_{(s)}$ for $s = 1, \ldots, |\boldsymbol{\tau}_2|$, in the order that maximizes the score statistic (\ref{TU}) in each step. In that case no more than $|\boldsymbol{\tau}_2|$ models will be fitted to reach the final model. In this paper we consider the penalties $\delta = 2$, as it is analogous to the \textit{Akaike} information criterion. It is also possible to use $\delta = \log N$ to have an analogous to the \textit{Bayesian} information criterion. 

\section{Results} \label{results}
In this section, we apply the McGLM for multivariate count data to 
analyse the data set presented in Section~\ref{dataset}. 
The second-moment assumptions of the McGLM require the specification 
of a linear predictor and a matrix linear predictor for each response 
variable. In this application, for composing the linear predictor 
we have three covariates \texttt{sex}, \texttt{method} and \texttt{alt} 
along with the time trend \texttt{month}. We considered interaction terms 
up to second order between the four main effects. 
The time trend was modelled as a polynomial of third and fourth degrees 
for \texttt{BD} and \texttt{OT}, respectively. 
Such choices were based on exploratory analysis and preliminary fits 
as we shall explain better in the Section~\ref{discussion}.
In all fitted models the number of hunter days (in logarithm scale) was 
used as an \texttt{offset}.

To specify the matrix linear predictor, we have the repeated measures
structures represented by the \texttt{Hunter} and \texttt{Hunter-Month} 
effects. The \texttt{Longitudinal} effect introduced by the observations 
taken at sequentially months and the three covariates, \texttt{sex}, 
\texttt{method} and \texttt{alt}. For the repeated measures effects
we assumed a compound symmetry (of ones) structure,see~(\ref{compsyme}).
The longitudinal effect was modelled using the inverse of Euclidean 
distances,see~(\ref{Eucli}). Finally, the covariates are included
in the covariance model in a linear mixed model fashion, see \ref{matlinear}. 
In this application for model parsimony and 
since we have only categorical covariates to compose the matrix linear
predictor, we considered only main effects. 

For clarity, consider a particular \texttt{Hunter} that represents the group 
structure described in the Section~\ref{model}. Furthermore, consider 
that we have four observations (two for the first month and two 
for the second month). Consider also for simplicity that we have the values 
of a covariate $\boldsymbol{e} = (e_1, e_2, e_3, e_4)$. In that case, the 
matrix linear predictor has the following form

\begin{eqnarray*}
\boldsymbol{\Omega}(\boldsymbol{\tau}) = \tau_0 \begin{bmatrix}
1 & 0 & 0 & 0\\ 
0 & 1 & 0 & 0\\ 
0 & 0 & 1 & 0\\ 
0 & 0 & 0 & 1
\end{bmatrix} +
\tau_1
\begin{bmatrix}
1 & 1 & 1 & 1 \\ 
1 & 1 & 1 & 1 \\ 
1 & 1 & 1 & 1 \\ 
1 & 1 & 1 & 1
\end{bmatrix} + 
\tau_2
\begin{bmatrix}
1 & 1 & 0 & 0 \\ 
1 & 1 & 0 & 0 \\ 
0 & 0 & 1 & 1 \\ 
0 & 0 & 1 & 1
\end{bmatrix} + \\
\tau_3
\begin{bmatrix}
0 & 0 & 1/d_{12} & 1/d_{12} \\ 
0 & 0 & 1/d_{12} & 1/d_{12} \\ 
1/d_{12} & 1/d_{12} & 0 & 0 \\ 
1/d_{12} & 1/d_{12} & 0 & 0
\end{bmatrix} +
\tau_4 
\begin{bmatrix}
e_1^2 & e_1 e_2  & e_1 e_3 & e_1 e_4 \\ 
e_1 e_2 & e_2^2  & e_2 e_3 & e_2 e_4 \\ 
e_1 e_3 & e_2 e_3 & e_3^2 & e_3 e_4 \\ 
e_1 e_34& e_2 e_4 & e_3 e_4 & e_4^2 
\end{bmatrix},
\end{eqnarray*}
where $\tau_0$ is the \texttt{intercept} of the covariance linear model.
The parameters $\tau_1$, $\tau_2$, $\tau_3$ and $\tau_4$ measure the 
\texttt{Hunter}, \texttt{Hunter-Month}, \texttt{Longitudinal} and
covariate effects, respectively.

We employed a stepwise procedure for selecting the components of the 
linear and matrix linear predictors. The SIC using penalty $\delta = 2$ 
and the Wald test were used in the forward and backward steps, respectively.
We defined as stop criterion SIC $> 0$, since in that case the penalty 
is larger than the score statistics.

Our strategy to select the final model consists of: 
i) select the components of the linear predictor for each response 
variable fixing the covariance structure assuming independent 
observations, i.e. $Z_0 = \mathrm{I}$.
ii) select the components of the matrix linear predictor for each 
response variable fixing the mean structure obtained in step (i).
iii) fit the multivariate model and 
iv) remove non-significant effects in both linear and matrix linear 
predictors if any.
In this application after fit the multivariate model all covariates 
selected to compose the linear and matrix linear predictors were significant.
Supplemenaty Tables S$1$ and S$2$ present the step-by-step procedure. 
Table \ref{Wald} presents the Wald statistics for the components of the 
selected linear predictor for each response variable obtained by fitting
the final multivariate model.
The selected matrix linear predictors were composed of a diagonal matrix 
(\texttt{Intercept}) combined with the \texttt{Hunter-Month}, 
\texttt{Method} and \texttt{Longitudinal} effects 
for \texttt{BD} and only the \texttt{Hunter-Month} effect for \texttt{OT}.

\begin{table}
\centering
\caption{Wald statistics ($\chi^2$), degrees of freedom (Df) and p-values 
for the components of the selected linear predictor for each response variable.}
\label{Wald}
\begin{tabular}{lccc|lccc} \hline
\multicolumn{4}{c}{\texttt{BD}}             & \multicolumn{4}{c}{\texttt{OT}}                     \\ \hline
Effects            & Df   & $\chi^2$  & p-value & Effects               & Df   & $\chi^2$   & p-value \\ \hline
\texttt{method}    & $1$  & $6.986$   & $0.008$  & \texttt{method}       & $1$  & $1.766$   & $0.183$   \\
\texttt{alt}       & $4$  & $138.262$ & $0.000$  & \texttt{alt}          & $4$  & $128.042$ & $0.000$   \\
\texttt{sex}       & $1$  & $247.843$ & $0.000$  & \texttt{sex}          & $1$  & $15.927$  & $0.000$  \\
\texttt{month}     & $3$  & $25.791$  & $0.000$  & \texttt{month}        & $4$  & $10.150$  & $0.038$   \\
\texttt{method:alt}& $4$  & $58.688$  & $0.000$  & \texttt{method:alt}   & $4$  & $26.455$  & $0.000$  \\
\texttt{alt:month} & $12$ & $43.898$  & $0.000$  & \texttt{alt:sex}      & $4$  & $13.238$  & $0.012$   \\ 
$-$                & $-$  & $-$      & $-$       & \texttt{alt:month}    & $16$ & $90.365$  & $0.000$ \\ \hline
\end{tabular}
\end{table}

The results in Table \ref{Wald} show that the \texttt{method} effect for the response 
variable \texttt{OT} was non-significant, but given its highly significant interaction 
with \texttt{alt} we opted to keep this effect in the model.
Table \ref{covariance} shows the estimates, standard errors (SE) and Z-statistics 
for the power and dispersion parameters for the final model. 

\begin{table}
\centering
\caption{Power and dispersion parameter estimates, standard errors (SE) and Z-statistics 
for the components of the selected matrix linear predictor for each response variable.}
\label{covariance}
\begin{tabular}{lccc|ccc} \hline
                      & \multicolumn{3}{c}{\texttt{BD}} & \multicolumn{3}{c}{\texttt{OT}}                     \\ \hline
Effects               & Estimate  & SE       &  Z-statistics &   Estimate &  SE      &  Z-statistics  \\ \hline
\texttt{power}        & $1.165$   & $0.115$  & $10.108$      & $1.453$    & $0.251$  & $5.777$ \\
\texttt{Intercept}    & $0.474$   & $0.142$  & $3.345$       & $0.686$    & $0.184$  & $3.737$  \\
\texttt{Hunter-Month} & $0.722$   & $0.151$  & $4.792$       & $0.294$    & $0.093$  & $3.163$  \\
\texttt{Method}       & $0.928$   & $0.258$  & $3.603$       & $-$        & $-$      & $-$       \\
\texttt{Longitudinal} & $-0.155$  & $0.0424$ & $-3.660$      & $-$        & $-$      & $-$      \\ \hline
\end{tabular}
\end{table}
The estimates of the power parameters suggest that the Neyman 
Type A ($p=1$), which indicates a zero inflation relative to the Poisson
distribution is a suitable choice for both response variables.
For the response variable \texttt{OT} the 
P{\'o}lya-Aeppli ($p=1.5$) can also be suggested.
The correlation between response variables was weak 
$-0.0532$ ($0.0287$) and not significantly different from $0$.

It is interesting to highlight that the $\boldsymbol{\Omega}$ matrix 
describes the part of the covariance structure that does not depend on 
the mean structure. Thus, it is interesting to interpret the parameters that
compose this matrix in terms of the correlation introduced by its components.
For example, the correlation introduced by the \texttt{Hunter-Month} 
effect is $0.604 (0.0594)$ and $0.299 (0.102)$ for \texttt{BD} and 
\texttt{OT}, respectively. These numbers are easily obtained by 
$\hat{\tau}_1/(\hat{\tau}_0 + \hat{\tau}_1)$. Similarly, 
the correlation between observations taken at the same 
hunter by the method snare is $0.652 (0.074)$. Note that, since the
\texttt{Hunter} effect was not significant the reference level is the
\texttt{Intercept} i.e. independence. Thus, we have no evidence of
dependence between observations taken at the same hunter by the method
firearm. Finally, the correlation introduced by the \texttt{Longitudinal} 
effect is $-0.487 (0.203)$ for lag equals $1$. 
The numbers in the brackets denote the standard error computed using 
the delta method.

\begin{figure}[htbp]
\centering
\includegraphics{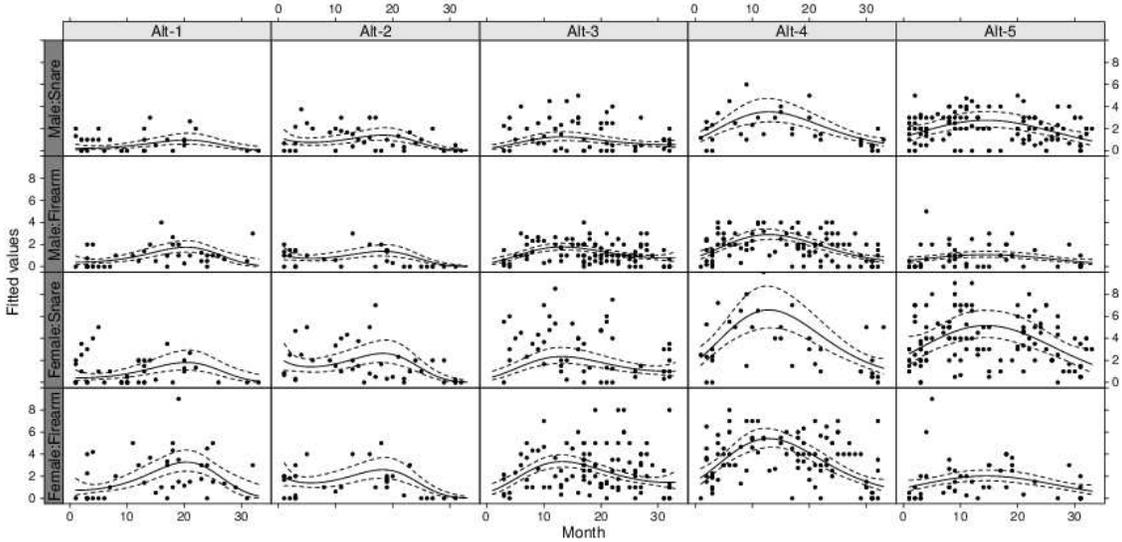}
\caption{Fitted values and $95\%$ confidence intervals by altitude, 
method of capture and sex for the response variable \texttt{BD}.}
\label{fig:bd}
\end{figure}

\begin{figure}[htbp]
\centering
\includegraphics{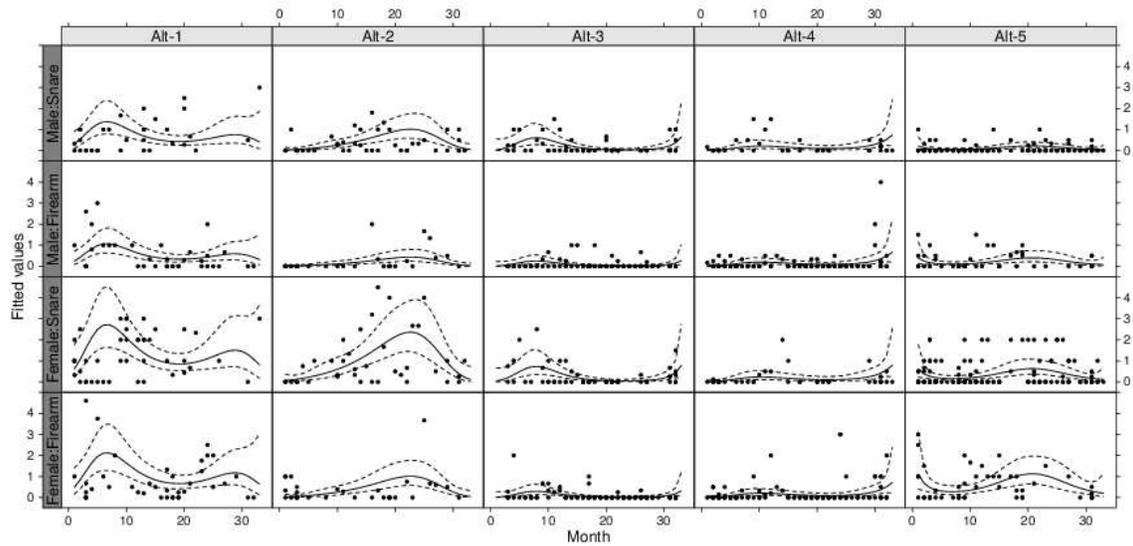}
\caption{Fitted values and $95\%$ confidence intervals by altitude, 
method of capture and sex for the response variable \texttt{OT}.}
\label{fig:ot}
\end{figure}

Figures \ref{fig:bd} and \ref{fig:ot} present the fitted values and 
$95\%$ confidence intervals for the response variables \texttt{BD} and 
\texttt{OT}, respectively. We plot the observed values divided by the
\texttt{offset} and the fitted values were computed fixing the 
\texttt{offset} equals $1$. Supplementary Tables S$3$ and S$4$
present the estimates and standard errors for the regression coefficients 
associated with the response variables \texttt{BD} and 
\texttt{OT}, respectively.

\setkeys{Gin}{width=0.99\textwidth}

\setkeys{Gin}{width=0.99\textwidth}

Figure \ref{fig:bd} shows that for all altitudes the number of
hunted blue duikers increases from the beginning to the middle of the 
data collection, when a clear decreases start with 
sensible differences in the threshold point among the levels of the 
covariate \texttt{alt}.
Altitudes $4$ and $5$ present the largest numbers of caught animals 
while altitudes $1$ and $2$ the smallest ones. 

Similar we have seen for \texttt{BD} Figure \ref{fig:ot} shows a clear 
time trend for the response variable \texttt{OT} in the altitudes 
$1$ and $2$. Altitudes $3$ and $4$ show a different pattern with a 
slightly increase at the end of the experiment.
Altitudes $1$ and $2$ present the largest numbers of other animals 
hunted by both methods and sexes. The smallest numbers appear in altitudes 
$3$ and $4$ using firearms. In general the number
of females hunted is bigger than males and the most effective method of
capture depends on the altitude.

It is important to highlight that despite of the differences in terms of 
altitudes, sexes and methods seem small in its magnitude judging by 
the results presented in Figures \ref{fig:bd} and \ref{fig:ot}. 
Such impression is due to the fact that, such results were obtained by 
fixing the number of hunter days (\texttt{offset}) equals $1$. 
Thus, the differences tend to be amplified while the number of hunter 
days increases. 
Furthermore, the regression coefficients associated with these effects
are in general significantly different from $0$ (see Tables \ref{Wald},
S$3$ and S$4$).

\section{Discussion} \label{discussion}

This section discusses the results presented in Section \ref{results}. 
The main data analysis goal was to determine if there was evidence of 
depletion in the population of blue duikers and other small animals 
based on data of hunted animals. 
To detect such a depletion effect, we included in the model a special 
term that represented the time trend for which we allowed a flexible 
functional form through a polynomial of degree three and four for 
the response variables \texttt{BD} and \texttt{OT}, respectively. 
To control other effects that were not of main interest, we included in 
the model the effects of covariates such as sex, method of capture and 
altitude. The irregular activity of the hunters introduces severely 
unbalanced repeated measures and longitudinal structures that were 
modelled through a matrix linear predictor composed of known matrices. 
Although these effects are not 
of main interest, they help us to understand the complex dynamics of 
hunting activity and provide us with insights of the general aspects of 
the population of the targeted taxa. In what follows we discuss the 
effect of all covariates.

The results presented in Section \ref{results} showed that for both 
response variables (\texttt{BD} and \texttt{OT}), 
methods (snare and firearm) and all altitudes,  the number of females 
hunted was larger than males.
Since hunters do not target any particular animal, this bias in sexes 
hunted could be a function of a greater hunting susceptibility of 
females or that there are more females in the population than males. 
With regards to the method of capture, our results showed that this 
covariate presents a highly significant interaction with the covariate
altitude. For the response variable \texttt{BD} the regression 
coefficients presented in the Supplementary Table S$3$, show that the method 
firearm is the most effective in altitude $1$, while the method snare is
the most effective in altitude $5$. For altitudes $2$ to $4$ the 
differences between the methods of capture are not significant.
Regarding the response variable \texttt{OT} the method snare is the
most effective in altitudes $2$ and $3$, while the method firearm is the
most effective in altitude $5$. In the altitudes $1$ and $4$ there is no
difference between the methods.

The covariate altitude reflected different hunting pressure at variable 
elevations in the study areas. Blue duikers may be overhunted in lower 
altitudes ($1$ and $2$) because of the proximity to human settlements, 
which increases hunting pressure. It may explain why the number of 
blue duikers is lower in altitudes $1$ and $2$. 
On the other hand, in altitudes $3$ to $5$ we presume that more animals 
are hunted because these areas are less exploited areas. 
The opposite situation appears for other small animals, this result
may indicate a depletion effect. Often, when the bigger animals 
(such as blue duikers in Bioko) are hunted out, which may be happening 
in altitudes $1$ and $2$, smaller ones tend to increase in numbers. 
This phenomenon is known as density compensation \citep{Fa:2009}.

While modelling the covariance structure we detected a significant 
effect of the covariate \texttt{Hunter-Month} for \texttt{BD} and \texttt{OT}. 
This effect is clearly due to the way that the data were collected and the 
arbitrary monthly aggregation. For the response variable \texttt{BD} 
in addition to the \texttt{Hunter-Month} effect, the longitudinal structure 
showed a significant negative effect. This result indicates that hunters 
may be affecting the prey population. Hence, some time is required for 
the population to recover and may indicate overexploitation of the hunted 
blue duikers population. 
A strong correlation between observations taken by the method snare was 
detected, but none appeared between observations taken by the method firearm. 
Such result is expected since the use of firearms to hunt is more 
effective when killing larger animals, so we would expect that the number of 
prey to decline with hunting effort with guns. 
This effect was detected by the longitudinal effect. 
On the other hand, because the method snare requires a much more 
continuous effort, the observations are more similar and consequently 
correlated along the study period. 
This mix of methods of capture could explain that months with a large 
number of animals hunted were followed by months with a smaller number 
of animals taken, explaining the negative longitudinal effect detected.

Finally, the time trend showed that for the response variable 
\texttt{BD} the number of hunted animals increases from the beginning 
to the middle of the data collection, followed by an intense decline 
after that. The maximum number of animals hunted appeared around the 
months $20$ and $14$ for altitudes $1$ to $2$ and $3$ to $5$, respectively. 
A possible explanation for this result could be that at the 
start of the study period the blue duiker population in the region were 
more numerous, but following intensive hunting the population starts 
to decrease and consequently the number of hunted animals also falls. 
Another explanation could be that there is interannual variation in 
numbers which may be related to changes in climate and by consequence 
productivity of the forest, but we have no additional data to confirm 
this hypothesis. The significant decline after the middle of the study 
period provides support for an overhunting effect. 

The temporal pattern detected for the response variable \texttt{OT} 
is more volatile mainly in altitudes $1$ and $2$, indicating that the 
number of \texttt{OT} animals hunted could have been affected by many 
factors, including the availability of other species as well as 
economic and climate conditions. This volatile pattern may also
explain the weak and non-significant correlation between \texttt{OT}
and \texttt{BD}.

Modelling the time trend through a polynomial function was a data-driven 
decision based on exploratory analysis and preliminary fits. 
The preliminary fits consisted of fitting models using B-splines basis
as implemented in the package \texttt{splines} for the \texttt{R}
statistical software. To select the number of degrees of freedom required
for the B-splines basis, we fitted models using different degrees of freedom
and check the significance of their regression coefficients using Wald test. 
Based on this procedure, we obtained that for the response variables \texttt{BT} 
and \texttt{OT} three and four degrees of freedom were enough to provide a 
suitable fit. Furthermore, based on the behaviour of the fitted values and
given the low number of degrees of freedom required by the B-spline basis,
we detected that a simple polynomial could provide a suitable fit.
Thus, we fitted the model changing the B-spline basis by polynomial of three
and four degrees of freedom for the response variables \texttt{BT} and 
\texttt{OT}, respectively.

We compared the fitted model with the one obtained by using the B-spline basis 
in terms of Gaussian pseudo-likelihood (GPL)~\citep{Carey:2011}. 
GPL is a measure similar to the log-likelihood value in the context of 
maximum likelihood estimation. Thus, larger values indicate better fit. 
The value of the GPL for the model presented in the 
Section~\ref{results} was $-4463.330$.
Similarly, the value of the GPL for the model fitted using
the B-splines basis was $-4462.270$. The GPL indicated that the fits 
are quite similar. Furthermore, we also compared the fitted values obtained from
both models that were virtually the same. Thus, we opted to present the model fitted 
using the polynomial. The advantage of the polynomial is that it is more familiar 
to applied researchers than the B-spline basis.

To provide more sources of evidence that the data support the model
presented in the Section~\ref{results} we fitted models using 
linear and quadratic time trends. 
The value of the GPL for the model fitted by using 
the linear trend was $-4572.300$.
Similarly, the value of the GPL for the model fitted by using the
quadratic trend was $-4477.670$.
Thus, we have clear evidences that the model presented in the 
Section~\ref{results} provides the best fit among the polynomial 
alternatives considered to describe the time trend.
Furthermore, the same conclusion is obtained when penalizing the
Gaussian pseudo log-likelihood with penalties compatible with the
Akaike and Kullback-Leibler information criterion~\citep{Bonat:2016a}.

\section{Concluding remarks} \label{conclusion}

We presented a flexible class of multivariate models for handling 
count data. The models were motivated by a data set
consisting of the number of blue duikers and other small animals
shot or snared by $52$ commercial hunters in Bioko Island, Equatorial Guinea.
The analysis of the data showed interesting features as overdispersion,
excess of zeroes and negatively correlated response variables, which in turn
allowed to show the flexibility of our models.

In our framework overdispersion and excess of zeroes are taken into 
account by means of a dispersion function. It is similar 
to a variance function in the context of generalized linear models. 
The dispersion function allows to specify models based only on 
second-moment assumptions and adopts an estimating function 
approach for parameter estimation and inference. 
The advantage of the estimating function approach is that the 
estimation procedure relies on a simple and efficient Newton 
scoring algorithm.
In this paper, we adopted the dispersion function associated with 
the Poisson-Tweedie distribution, since important discrete distributions 
as the Neyman Type A, negative binomial and Poisson-inverse Gaussian 
appear as special cases. 

The marginal covariance structure within response variables is
specified by means of a matrix linear predictor composed of known 
matrices. This specification easily deals with the combination
of unbalanced repeated measures and longitudinal structures 
as well as the effects of the covariates in a linear mixed model 
fashion. The flexibility of this structure comes with the issue 
to select its components. In this paper, we extended the SIC to 
guide the selection of the matrix linear predictor components.
The great advantage of the SIC is its simplicity.
Since the SIC is based on the score statistics it can be computed 
without actually fitting all the candidate models. 

The strategy employed in this paper for selecting the components of the
linear and matrix linear predictors consisted of combining the SIC and 
Wald statistics in a stepwise procedure applied independently for the mean
and covariance structures.
In the first step, we selected the components of the linear predictor
for each response variable assuming independent observations. 
In fact, in this step we are purposely ignoring the correlation 
between and within response variables.
It is well known that in the presence of correlation the standard errors 
associated with the regression parameters are underestimated. 
In this way, we avoid to remove important covariates of the analysis.
In the second step, we fixed the linear predictor as obtained 
in the first step and selected the components of the matrix linear 
predictor. As the linear predictor potentially contains all
significant covariates, we avoid that missing covariates 
affect the selection of the matrix linear predictor components. 
In the last step, we fit the multivariate model and remove any
non-significant effect.

Finally, the joint covariance matrix is specified by using the generalized
Kronecker product. This specification combined with the possibility to 
estimate the power parameter for each marginal response variable allow 
our models easily deal with negatively correlated and unequal marginal 
response variables, overcoming the main limitations of the multivariate 
Poisson and negative binomial models. 

The main limitation of the models presented in this paper is the general
lack of algorithms for simulation. Recent work of \citet{Baccini:2015}
discussed the problems involving the simulation of univariate
Poisson-Tweedie distributions. The related topic of simulation of the
multivariate Tweedie distributions was addressed recently by \citet{Cuenin:2015},
but the extension to multivariate Poisson-Tweedie distributions specified by
general covariance structures in high dimension, as used in this paper, still
requires further theoretical and computational developments.

\section*{Supplement material}
Dataset and R code for the analysis are available at the paper companion page at
\texttt{http://www.leg.ufpr.br/doku.php/publications:papercompanions:hunting\\bioko2016}.
The authors thank Professors Elias Teixeira Krainski, Walmes Marques Zeviani, Fernando Poul Mayer and Paulo Justianiano Ribeiro Jr for their comments and suggestions that substantially improve the article.
The first author is supported by CAPES (Coordena\c{c}\~ao de Aperfei\c{c}oamento de Pessoal de N\'ivel Superior)-Brazil.

\bibliographystyle{dcu}
\bibliography{Bonat2016}

\end{document}